# Ultrafast terahertz conductivity in epitaxial graphene nanoribbons: an interplay between photoexcited and secondary hot carriers

*Arvind Singh[1], Hynek Němec[1], Jan Kunc[2], and Petr Kužel[1*]*

*[1] FZU - Institute of Physics of the Czech Academy of Sciences, Na Slovance 2*
*18221 Prague 8, Czech Republic*

*[2] Faculty of Mathematics and Physics, Charles University, Ke Karlovu 3,*
*12116 Prague 2, Czech Republic*

[*] kuzelp@fzu.cz

**Abstract:** Optical pump-terahertz probe spectroscopy has been used to investigate ultrafast photo-induced charge carrier transport in 3.4 μm wide graphene ribbons upon scaling the optical pump intensity. For low pump fluences, the deposited pump energy is rapidly redistributed through carrier-carrier scattering, producing secondary hot carriers: the picosecond THz photoconductivity then acquires a negative sign and scales linearly with an increasing pump fluence. At higher fluences, there are not enough equilibrium carriers able to accept the deposited energy, directly generated (excess) carriers start to contribute significantly to the photoconductivity with a positive sign leading to its saturation behavior. This leads to a non-monotonic variation of the carrier mobility and plasmonic resonance frequency as a function of the pump fluence and, at high fluences, to a balance between a decreasing carrier scattering time and an increasing Drude weight. In addition, a weak carrier localization observed for the polarization parallel to the ribbons at low pump fluences is progressively lifted upon increasing the pump fluence as a result of the rise of initial carrier temperature.

**Keywords:** epitaxial graphene, ultrafast photoconductivity, ultrafast carrier dynamics, terahertz

## 1. Introduction

Hot carriers in graphene exhibit fascinating physical properties that are distinctly different from the response of conventional semiconductors or metals. [1–5] This distinction arises predominantly due to the remarkably low electronic heat capacity of massless Dirac fermions, characterized by a linear electronic dispersion with strong carrier-carrier interactions and relatively weak coupling between the electronic and phononic systems. [6–11] Notably, photo-carriers can be easily excited at energies well above the Fermi level, emphasizing the subsequent carrier relaxation mechanisms pivotal for the functionality of opto-electronic devices. In addition, graphene exhibits the strongest nonlinear THz response to date, which is attributed to hot carrier dynamics, [12–15] enabling, e.g., efficient high harmonic generation and paving the way for practical graphene-based applications in ultrafast (opto-)electronics operating at THz frequencies.

The photoinduced carrier dynamics in graphene have been investigated intensively since 2008. [1,16–19] Various ultrafast measurements have been conducted to explore non-equilibrium effects, including carrier multiplication, population inversion, and charge transport under different experimental conditions. [3,20–25] The carrier relaxation in graphene is highly influenced by its linear electronic band structure and depends also on the doping density. It involves a series of ultrafast processes, including thermalization within 10 – 100 fs, cooling within 0.1 – 10 ps, and recombination occurring on picosecond time scales. [3,11,17,24]

Notably, saturation of various properties (including the carrier temperature or the carrier cooling rate) with an increasing excitation fluence has been attributed to hot phonon effects, which facilitate both free carrier absorption and re-absorption of hot optical phonons. [3,19–22] Moreover, almost all fabrication methods produce unavoidable morphological inhomogeneities that significantly affect the electronic transport. [26–29] In particular, structural defects like grain boundaries, vacancies, and wrinkles act as potential barriers which decrease the carrier mobility. [27,30,31] Further effects stem from an interaction with the substrate: for example, epitaxial graphene growth on SiC may introduce strain, doping, and charge traps which enhance the carrier localization. [32–34] Recently, nanoscale carrier confinement has been directly observed in polycrystalline CVD graphene via THz time-domain spectroscopy and attributed to a strong backscattering due to the grain boundaries. [35]

In this study, we provide a more quantitative analysis of the nonlinear scaling of terahertz conductivity response utilizing optical pump—terahertz probe spectroscopy across a broad range of optical excitation fluences. For this purpose, we probe hot-carrier plasmon modes excited in a system of graphene ribbons, which provide quite a rich and precise information about the carrier response. We demonstrate that this nonlinear dependence is not a result of the Pauli-blocking of interband transitions, but it rather arises from an interplay between directly photoexcited carriers (dominating at high fluences) and secondary carriers heated by the carrier-carrier scattering (dominating at low fluences). This leads to a non-monotonic variation of the carrier mobility and plasmonic resonance versus the pump fluence. We also quantitatively distinguish the role of the scattering time, carrier temperature, chemical potential and carrier mobility in the high pump fluence regime. A careful analysis of the measured conductivity spectra shows clear signatures of a weak carrier localization, which is attributed to the contamination during the lithography. This carrier confinement occurs only at low pump fluences, whereas at high fluences, the carriers exhibit Drude-type conduction since the increase in the initial carrier temperature allows the charges to pass over the confining potential barriers.

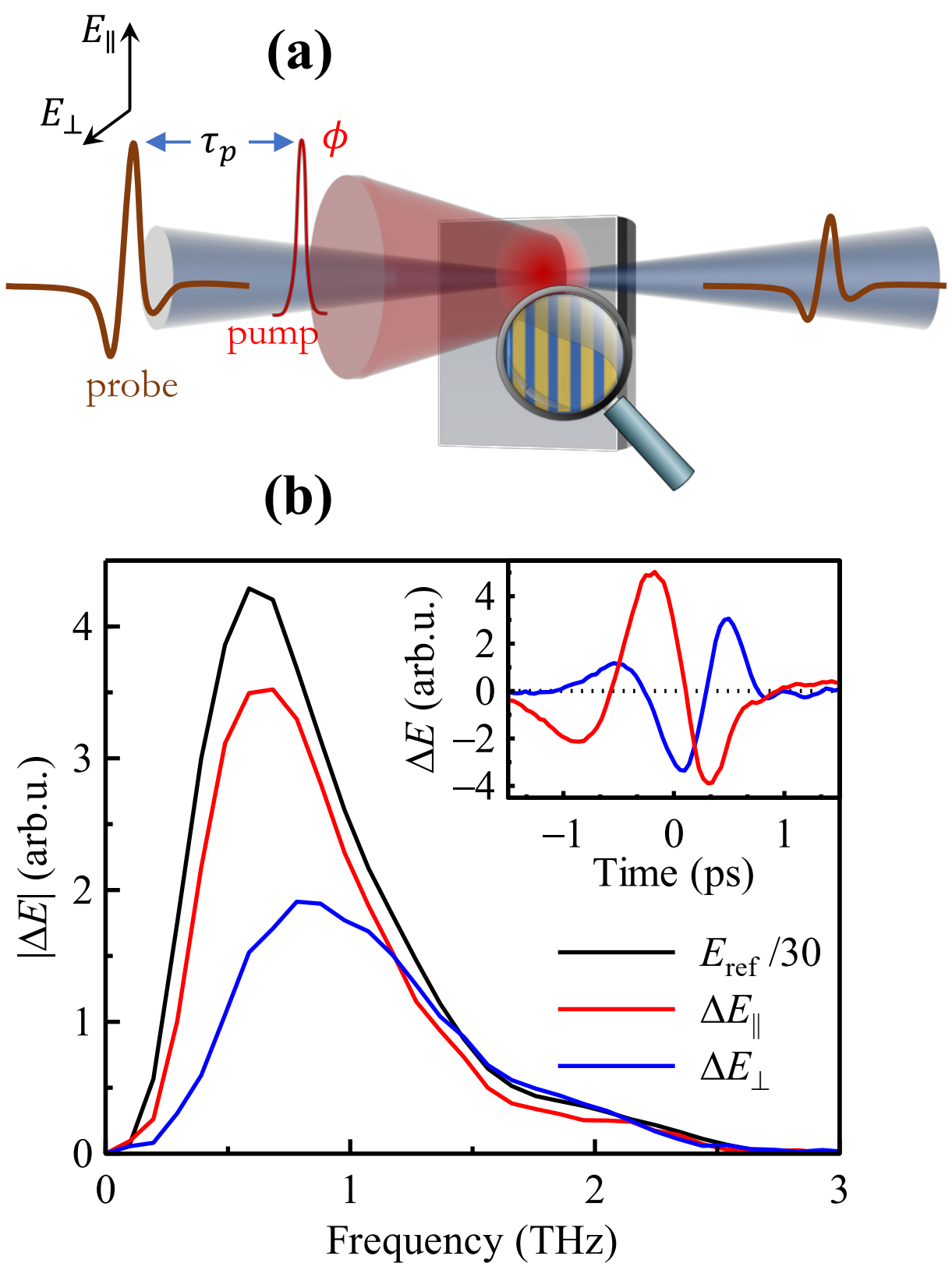

**Figure 1:** (a) Experimental scheme for optical pump – THz probe spectroscopy. Measurements are performed in a collinear geometry with THz electric field polarization both parallel and perpendicular to the graphene ribbons. (b) Examples of differential THz signals measured for both THz polarizations under the same conditions at 2 ps after photoexcitation at $2\times10^{12}$ photons/cm$^2$. Main plot: frequency domain amplitudes (including also the spectrum $E_{ref}$ transmitted through unexcited bare substrate); inset: differential time domain waveforms.

## 2. Experimental part

The graphene ribbons studied here are prepared in three steps. Initially, a graphene layer is epitaxially grown on 6H-SiC substrate by thermal decomposition technique; [36] subsequently, hydrogen intercalation is performed to decouple the graphene layer from the substrate (so called quasi-free standing single layer graphene is formed). [37] Finally, a large area ribbon array is fabricated using an electron beam lithography technique with a graphene ribbon width $w = 3400$ nm and a gap between the ribbons $g = 500$ nm (i.e., a period of the array $L = 3900$ nm). [38]

Figure 1(a) shows the experimental scheme which is used to measure the transient transmission of the epitaxial graphene ribbon sample. The optical pulses (time duration of 40 fs, wavelength of 800 nm and a repetition rate of 5 kHz) are delivered from a Ti:sapphire amplifier (Spitfire ACE, Spectra Physics). They are used for the sample photoexcitation at 800 nm and also for THz emission and detection ensured by a pair of (110) oriented 1 mm thick ZnTe crystals. The optical pump and terahertz probe pulse are collinearly incident on the sample surface covered by a 4 mm aperture. The pump beam is expanded in space to a spot with FWHM exceeding 7 mm to excite the sample homogeneously within the aperture.

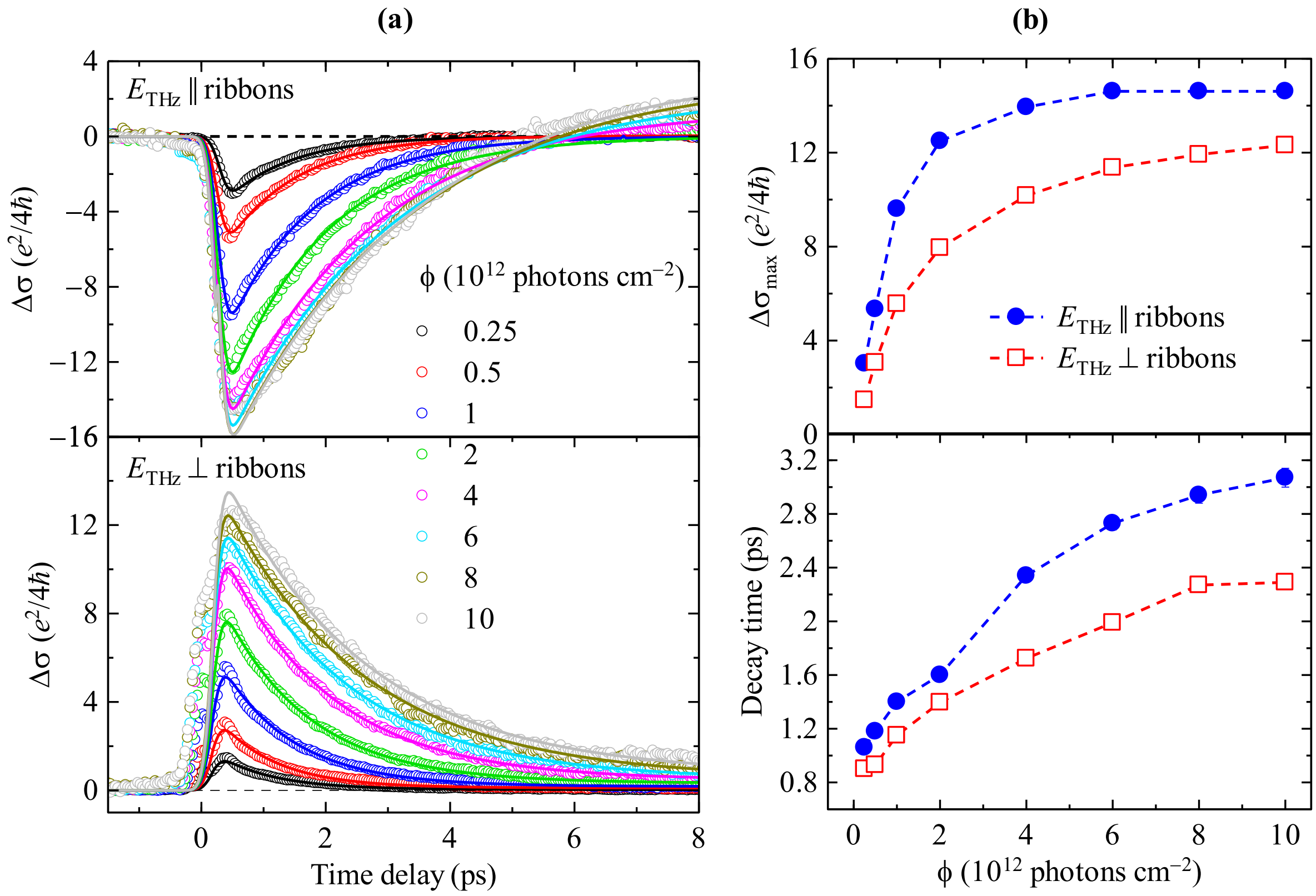

**Figure 2:** (a) Spectrally averaged THz photoconductivity dynamics of graphene ribbons for several absorbed pump fluences in parallel (top) and perpendicular (bottom) configurations of probing THz electric field. Solid lines in the kinetic signal are fits using a mono-exponential decaying function convolved with the Gaussian profile representing the pump pulse, Eq. (2). (b) Parameters extracted from the fit: Photoconductivity amplitude (top) and decay time (bottom) as a function of the absorbed pump fluence $\phi$.

## 3. Results and discussion

Figure 2(a) shows the picosecond dynamics of the average transient THz photoconductivity for both polarizations of the probing THz field with respect to the graphene ribbons. The curves (so called pump-probe scans) are measured at the absolute maximum of the differential waveform $\Delta E$ as a function of the pump-probe delay. The differential transmission $\Delta E$ is directly related to the (spectrally averaged) photoconductivity; in the thin film approximation and small signal limit ($\Delta E \ll E_0$) we obtain [39]

$$\Delta\sigma = -\frac{1+N_s}{z_0}\frac{\Delta E}{E_0} \tag{1}$$

where $E_0$ is a reference signal (transmitted THz field with the pump beam off), $N_s$ ($\approx 3.13$) is the THz refractive index of the SiC substrate and $z_0$ is the vacuum wave impedance. The small signal limit is well satisfied in our experiments since $\Delta E/E_0 < 10\%$ even for the highest pump fluence used. A nonlinear scaling of the dynamics versus the pump photon fluence absorbed in the graphene layer can be clearly observed in Fig. 2. Note also the different signs of the two curves (for the parallel and perpendicular configuration), which have been discussed in [38] and which will be briefly revisited later in this paper in relation with the measured time-resolved THz spectra. The temporal evolution of the pump-induced THz transient dynamics is analyzed simply by a numerical fit of the experimental data using the following response function:

$$\Delta\sigma(t) = \Delta\sigma_{\mathrm{SiC}}(t) + \left[H(t-t_0) \times \Delta\sigma_{\max}\exp\left(\frac{t_0-t}{\tau_c}\right)\right] \otimes G(t) \tag{2}$$

where, $H(t)$ is the Heaviside function and $G(t)$ is an instrumental Gaussian time profile; the symbol $\otimes$ represents the convolution operation. The parameter $\Delta\sigma_{\mathrm{SiC}}$ accounts for the substrate contribution, while $\Delta\sigma_{\max}$ and $\tau_c$ characterize the amplitude and decay time of the transient response of the graphene film, respectively.

Figure 2(b) presents the fitting outcomes: the peak graphene photoconductivity in time $\Delta\sigma_{\max}$, and the decay time, $\tau_c$, as a function of absorbed pump photon fluence. At low fluences the peak photoconductivity scales linearly and then saturates at higher excitation fluences for both polarization configurations. The lifetime progressively increases with increasing excitation fluence.

Upon photoexcitation, primary electron-hole pairs are generated, initiating a sequence of carrier-carrier scattering events. This process involves the transfer of energy deposited by the pump beam (and initially acquired by the excess carriers) towards the conduction carriers which lie close to the Fermi level. These (originally equilibrium) carriers are then promoted to higher energies, thus creating a distribution of secondary hot carriers.[1] At lower fluences the concentration of excess carriers is low and the transient THz signal is essentially due to an increase in the effective temperature of the carrier distribution (i.e., due to the secondary hot carriers). In this regime, as the absorbed photon density increases, the observed increase in the negative photoconductivity signal can be attributed to a higher temperature of the hot carrier distribution. In other words, the change in the photoconductivity is directly linked to the secondary hot-carrier excitation process and the decay time reflects the carrier cooling. At higher fluences the concentration of excess carriers increases; on the contrary, since the density of secondary hot carriers is limited by the doping level, their contribution to the photoconductivity becomes less important. An interplay of the two contributions then leads to the photoconductivity saturation. The energy transferred to the carriers significantly raises the electronic temperature and the hot carriers lose their energy primarily through the emission of optical phonons. At high fluences this leads to a fast increase of the optical phonon population; however, these phonons cannot decay into acoustic phonons fast enough, which leads to the so-called hot phonon bottleneck effect. [9] A part of the phonon energy is then scattered back to the electronic system; consequently, the energy relaxation of the electronic system slows down and so does the observed carrier relaxation time in Fig. 2b. These non-trivial results will be quantitatively demonstrated below upon a detailed analysis of the transient photoconductivity spectra. It is worth noting that, regardless of the sign of the photoconductivity for the two THz probing polarizations observed in Fig. 2(a), the fluence dependence of the extracted parameters follows a similar trend (with a minor renormalization caused by the spectral shift of the plasmonic response in the perpendicular configuration), thereby validating the explanation discussed above for both cases.

For a more quantitative analysis of the photo-induced THz response of graphene ribbons, we measured transient THz waveforms 2 ps after photoexcitation for a set of the absorbed pump fluences (the same set of fluences was used for the two polarizations) and we calculated the photoconductivity spectra using Eq. (1), see Fig. 3. For the parallel configuration, the real part of the photoconductivity is negative over the whole accessible frequency range; this is due to the intra-band conductivity reduction due to the heating of carriers and it correlates with the enhanced transparency signal observed in Fig. 2(a), top panel. For the perpendicular configuration, one observes a zero crossing in $\Delta\sigma_{\mathrm{Re}}$, which is caused by a spectral shift of the plasmonic response upon photoexcitation. On the first glance, this observation might look inconsistent with the positive contribution to the spectral average observed in Fig. 2(a), bottom panel. However, note that the THz probing pulses exhibit the maximum power roughly in the spectral interval of 0.4–1 THz, see Fig. 1(b), i.e., the sample response in this particular range determines the global aspect of the transient time-domain waveforms, which then present a dominating negative half-cycle [blue curve in the inset of Fig. 1(b)] corresponding to a positive average photoconductivity in agreement with Eq. (1).

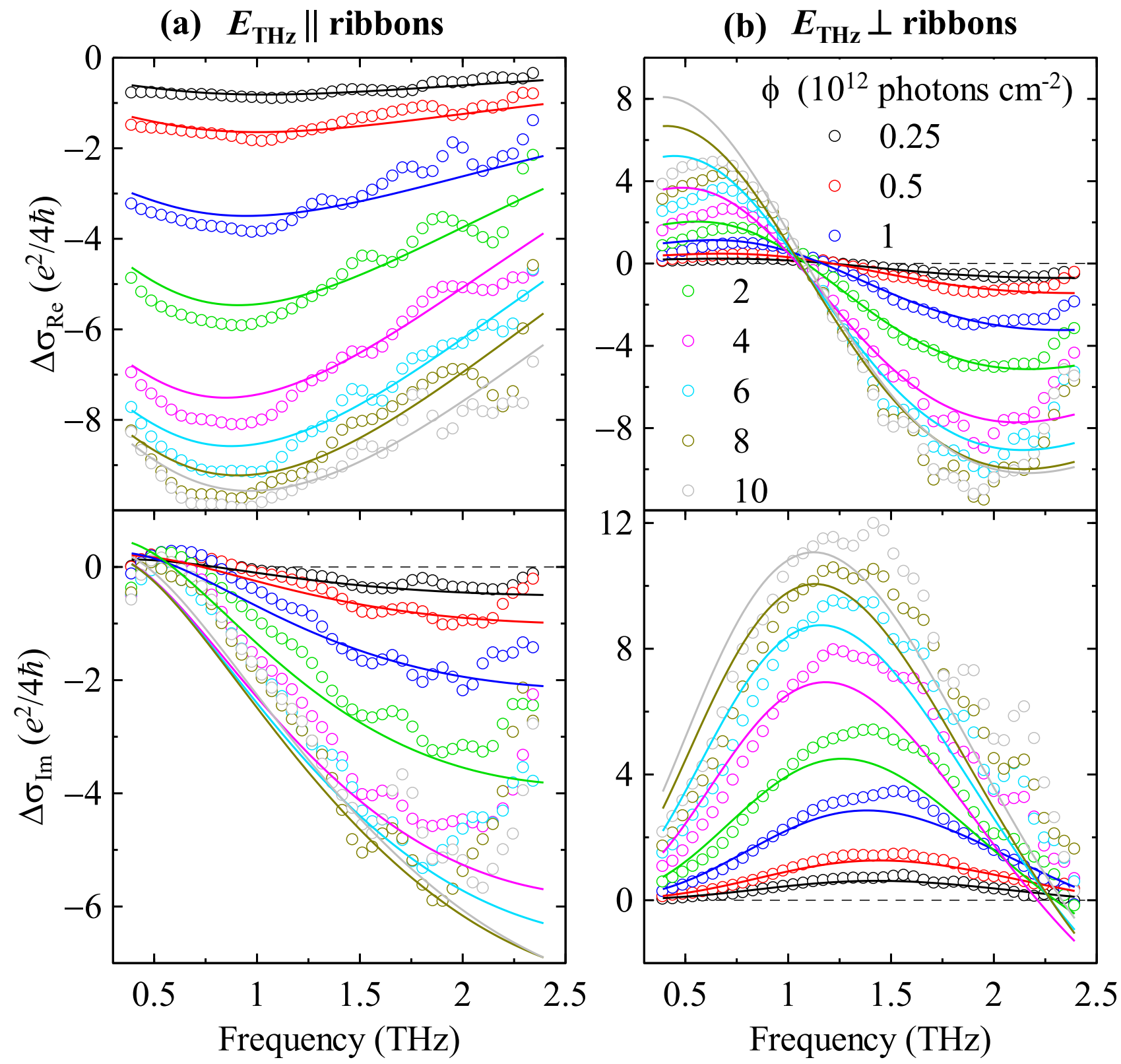

**Figure 3:** Transient sheet conductivity spectra as a function of the absorbed pump fluence for the pump-probe delay of 2 ps after photoexcitation in parallel (a) and perpendicular (b) configurations. Symbols: experimental points, lines: with the model defined by Eqs. (3)–(8).

The ultrafast photoconductivity spectra are fitted using the following model:

$$\Delta\sigma(\omega,\phi) = \sigma^{e}_{\perp,\parallel}(\omega,\phi) - \sigma_{\perp,\parallel}(\omega) + \Delta\sigma_{\text{SiC}}(\omega,\phi), \tag{3}$$

which involves the graphene sheet photoconductivity (the difference between the sheet conductivity of the excited state $\sigma^{e}_{\perp,\parallel}$ and of the ground state $\sigma_{\perp,\parallel}$) and also a weak photo-induced response from the SiC substrate $\Delta\sigma_{\text{SiC}}$; $\phi$ stands for the absorbed fluence of the pump pulse. The coupling of the electromagnetic radiation depends on the orientation of the structure: sheet conductivities in the excited and ground state in our sample are described by the Lorentz model for the perpendicular configuration [38,40] and by the modified Drude-Smith response for the parallel configuration [38]. They can be written, respectively, as

$$\sigma_{\perp}(\omega) = \frac{w}{L}\frac{D}{\pi}\frac{i\omega}{\omega^2 - \omega_0^2 + \dfrac{i\omega}{\tau_{s,\perp}}} \tag{4}$$

$$\sigma_{\parallel}(\omega) = \frac{w}{L}\frac{D}{\pi}\frac{\tau_{s,\parallel}}{1 - i\omega\tau_{s,\parallel}}\left(1 + \frac{c}{1 - \dfrac{i\omega d^2}{5\, v_F^2 \tau_{s,\parallel}}}\right) \tag{5}$$

where $w/L = 87\%$ is the filling fraction of the graphene component. The modified Drude-Smith model [41] accounts for an observed weak localization of carriers in the parallel geometry: $c$ describes the degree of localization (0 = no localization, –1 = full localization) and $d$ stands for the characteristic dimension of the confinement. The response is controlled by the Drude weight $D$ and the geometry of the ribbons, both determining the plasmonic resonance frequency $\omega_0$:

$$D = \frac{2e^2}{\hbar^2} k_B T_\mathrm{c} \ln\left[2\cosh\left(\frac{\mu(T_\mathrm{c})}{2\,k_B T_\mathrm{c}}\right)\right], \tag{6}$$

$$\omega_0 = \sqrt{\frac{Dw}{\varepsilon_0(1+\varepsilon_\mathrm{SiC})L^2 \ln\left(\sec\left(\frac{\pi w}{2L}\right)\right)}}. \tag{7}$$

In the ground state, the carrier temperature $T_\mathrm{c}$ is equal to 300 K. In the excited state, the carrier temperature is generally elevated. Furthermore, with an increasing pump fluence, the carrier density can be significantly modified alongside with the carrier temperature and momentum scattering time. This results in a small correction in the Fermi energy due to excess carriers resulting from the optical excitation:

$$E_F(\tau_p) = \sqrt{E_{F,SS}^2 + q(\tau_p)\phi\pi\hbar^2 v_F^2}\,, \tag{8}$$

where $E_{F,SS}$ is the steady-state (SS) Fermi energy and $q$ is a proportionality factor between 0 and 1 providing the fraction of non-recombined photocarriers at the time delay $\tau_p$. The chemical potential can be evaluated from the energy conservation law [42]

$$\frac{1}{2}\left(\frac{E_F}{k_B T_c}\right)^2 = F_1\left(\frac{\mu}{k_B T_c}\right) - F_1\left(-\frac{\mu}{k_B T_c}\right), \tag{9}$$

where $F_1$ is the first order Fermi-Dirac integral. Unfortunately, Eq. (9) yields a strong correlation between the photoexcited carrier density (determined by $q$) and temperature ($T_c$). Since our measurements were performed 2 ps after photoexcitation, we neglect in the first approximation the recombination during this short period, and we fix the photo-carrier fraction to $q = 1$. We performed a global fitting following the strategy described in detail in our previous publication [38]. First a global fit was performed on the ensemble of data in Fig. 3(b) for the perpendicular geometry. This yielded the steady-state Fermi energy value, as a global parameter, $E_{F,SS} = 310$ meV and the data shown in Fig. 4(a,c) represented by open red rectangles. The spectra in the parallel orientation, Fig. 3(a), were fitted using the carrier temperature values obtained from the fits in the perpendicular geometry leaving just the carrier scattering time and the confinement parameter $c$ as free parameters (independently fit for each fluence), along with the confinement length $d \approx 250$ nm (global parameter for all spectra); the results are represented by closed blue circles in Fig. 4(a,b). All the fit results obtained for low fluences are fully compatible with our previously published values obtained during a study of sub-picosecond and picosecond evolution of the system at a low fluence [38].

Our ad hoc choice of $q = 1$ is, of course, a crude approximation. For pump fluences lower than $2 \times 10^{12}$ photons cm$^{-2}$ the $q$-term in (8) is small (its contribution to the Fermi energy is at most $\sim 13\%$) and its value does not influence significantly the results of the fit. For higher pump fluences the importance of the choice of the value of $q$ becomes more important. We performed a global test fit of all the data for $q = 0.5$ and, based on the results, we can conclude that, even at the highest fluence, the difference in the resulting carrier temperature between the cases of $q = 1$ and $q = 0.5$ remains within the error bars shown in Fig. 4(a,c).

The evolutions of all the directly fitted parameters (scattering times, carrier temperature and confinement parameter) are shown in Fig. 4(a–c). The chemical potential, Drude weight, carrier mobility and plasmon resonance frequency are calculated from these fitted values and they are shown in Fig. 4(d–g).

Let us mention that the determined steady-state value of the Fermi energy corresponds to the free carrier concentration of $\sim 7 \times 10^{12}$ cm$^{-2}$. After photoexcitation, these carriers (or their part) will form the secondary (hot) carriers. The behavior of the investigated system can be divided into several regimes. For the low fluence regime ($\phi \lesssim 1 \times 10^{12}$ photons cm$^{-2}$), the carrier temperature almost linearly increases with the fluence since the secondary hot carrier generation is efficient and, consequently, the chemical potential and the Drude weight decrease rapidly [Fig. 4(d,e)]. As a consequence, and despite of a small decrease in the scattering time [Fig.(4a)], a small increase in the mobility $\eta = \tau_s e_0 v_F^2 / \mu(T_c)$ is observed with an increasing pump fluence. The plasmon frequency exhibits a red shift. The carrier temperature reaches $\sim 1400$ K and its growth with increasing fluence is then significantly reduced. We note that this temperature corresponds to the energy of optical phonons in SiC (120 meV), indicating that the energy of carriers exceeding this value can be dissipated into the substrate through substrate phonons.

Conversely, in the high fluence regime ($\phi \gtrsim 4 \times 10^{12}$ photons cm$^{-2}$), the generation of secondary hot carriers is much less efficient, since there is a limited number of equilibrium conduction carriers. As a result, the contribution of newly excited (excess) carriers takes over and this leads to an increase in the Drude weight, which compensates the observed decrease in the scattering time in Fig. 4(a). This balance accounts for the observed saturation of the photoconductivity amplitude [Fig. 2(b), top panel]. In this regime the decay time extends, and its growth does not saturate at the highest fluences [Fig. 2(b), bottom panel]; this is due to the above mentioned hot phonon bottleneck in the graphene ribbons, [9] which efficiently slow down the hot carrier relaxation. In this regime, the significant decrease in the carrier scattering time alongside with a rise in the chemical potential, lead to the observed decrease of the carrier mobility by up to factor of 4 (still under the assumption of $q = 1$, i.e., that all photocarriers remain in the conduction band till 2 ps after photoexcitation). The observed variation of the Drude weight leads to a complex behavior of the plasmon resonance; indeed, the red shift observed at low fluences is followed by a significant blueshift at high fluences due to the presence of excess carriers.

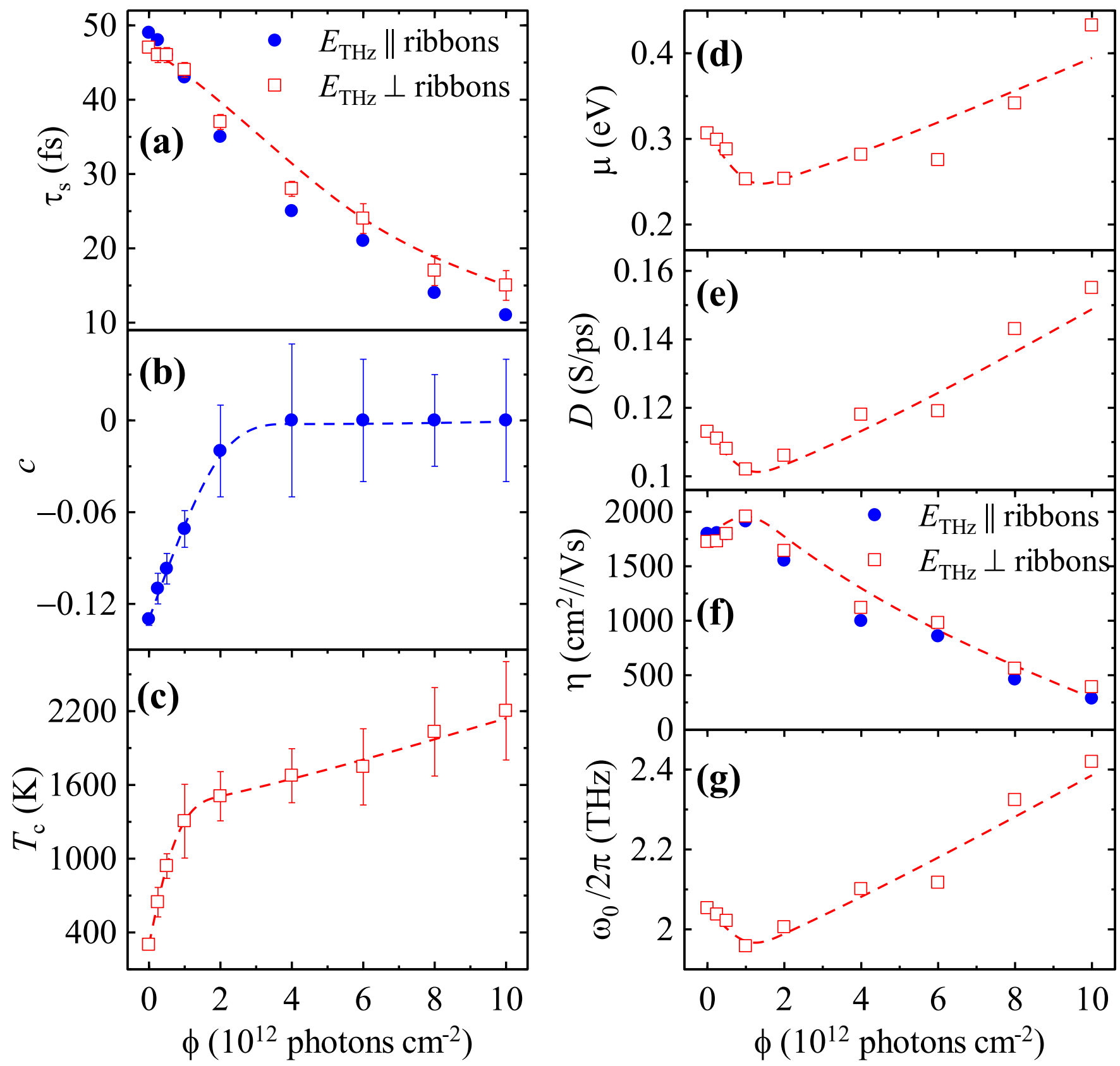

**Figure 4:** Evolution of the directly fitted parameters with the absorbed pump fluence: (a) scattering times $\tau_s$, (b) localization parameter $c$, and (c) carrier temperature $T_c$. Evolution of the calculated transport parameters and plasmonic frequency from the fit parameters: (d) chemical potential $\mu$, (e) Drude weight $D$, (f) mobility $\eta$, and (g) plasmonic frequency $\omega_0$. Red symbols are obtained from the spectra for $E_{THz} \perp$ ribbons and blue symbols are obtained from the spectra for $E_{THz} \parallel$ ribbons. Dashed lines are guides for the eyes.

The significant decrease of the scattering time excludes the state filling effects of pump photons in the conduction band. A weak localization of carriers ($c < 0$) was previously reported in this sample and its origins were discussed [38]; now it becomes clear that the localization is progressively lifted and it completely disappears ($c \to 0$) for $\phi \gtrsim 4 \times 10^{12}$ photons $\mathrm{cm}^{-2}$, see Fig. 4(b). This confirms that the localization is controlled by low energy barriers that are overcome by the hot carriers owing to their increased kinetic energy at elevated carrier temperatures. One can thus exclude the possibility of scarce defects consisting of borders with semi-infinite energy barriers.

## 4. Conclusion

We observed that increasing the fluence of optical excitation pulses drives the graphene into distinct regimes. (i) At low fluences, $\phi \lesssim 10^{12}$ photons/cm$^2$, an increase of the carrier temperature and a decrease of the chemical potential are observed; the charges undergo a weak confinement. (ii) In a transition (intermediate) regime, $10^{12}$ photons/cm$^2$ $\lesssim \phi \lesssim 4 \times 10^{12}$ photons/cm$^2$, a minimum of the chemical potential and a maximum of the mobility is observed due to an interplay of excess carriers and the equilibrium conduction carriers heated by the pump-deposited energy. (iii) The high fluence regime, $\phi \gtrsim 4 \times 10^{12}$ photons/cm$^2$, where the excess carriers dominate, is characterized by a saturation of the carrier temperature and conductivity.

**Acknowledgment**

We acknowledge the financial support by Czech Science Foundation (grant number 24-10331S) and by European Union and the Czech Ministry of Education, Youth and Sports (Project TERAFIT - CZ.02.01.01/00/22_008/0004594). Finally, CzechNanoLab project LM2023051 funded by MEYS CR is acknowledged for the financial support of the sample fabrication at CEITEC Nano Research Infrastructure.